\documentclass[aps,pra,preprint,superscriptaddress,tightenlines]{revtex4}
\usepackage{graphicx} 
\usepackage{amsmath}
 
\newcommand{\beq}{\begin{equation}}
\newcommand{\enq}{\end{equation}}

\begin{document}

\title{Longitudinal sound mode of a Bose-Einstein condensate 
in an optical lattice}
\author{J.-P. Martikainen}
\email{J.P.J.Martikainen@phys.uu.nl}
\author{H. T. C. Stoof}
\email{stoof@phys.uu.nl}
\affiliation{Institute for Theoretical Physics, Utrecht University, 
Leuvenlaan 4, 3584 CE Utrecht, The Netherlands}
\date{\today}
\begin{abstract}
We examine the effect of the 
transverse breathing mode on the longitudinal sound 
propagation in a Bose-Einstein condensate in
a one-dimensional optical lattice. In particular, we discuss
how the coupling with the transverse breathing mode
influences the sound velocity in an optical lattice.
Using a variational approach  we calculate the
dispersion relations for the longitudinal sound mode
and the transverse breathing mode analytically and
find that the shift in the sound velocity
from the uncoupled result can be large enough to be 
experimentally relevant. We also find that the effective 
mass of the transverse breathing mode is affected 
considerably by the coupling to longitudinal sound.
\end{abstract}
%\narrowtext
\maketitle

\section{Introduction}
\label{sec:intro}

%Must be gap-less in accord with the
%Hugenholtz-Pines theorem~\cite{Hugenholtz1959a,Hohenberg1965a}.
%Boliubov approximation is the simplest and works
%in $T=0$. Relation of the $k=0$ sound to the Goldstone-mode
%resulting from $U(1)$ symmetry breaking.
Repulsive contact interactions between atoms in a
dilute Bose gas 
cause a qualitative change in the low-energy
excitations of the system. In particular, the noninteracting single-particle
dispersion equal to $\hbar^2k^2/2m$ is replaced with the sound dispersion
$ck$ that is linear in the momentum~\cite{Bogoliubov1947a}. 
The existence of this sound
mode is important as it allows for the possibility of
superfluid flow~\cite{Nozieres1990a}. Sound propagation in
a trapped Bose-Einstein condensate was studied 
experimentally by Andrews 
{\it et al.}~\cite{Andrews1997b,Andrews1998a} and 
Stamper-Kurn {\it et al.}~\cite{StamperKurn1999a} 
in the regime when the condensate 
can be considered as being homogeneous in
one direction. The experimental observations
were in good agreement with theoretical 
predictions in that case~\cite{Zaremba1998a,Kavoulakis1998a,Stringari1998a}.

Bose-Einstein condensates can also be placed in an optical 
lattice. The periodic structure of the lattice has enabled
studies of Bloch oscillations~\cite{Cristiani2002a},
number squeezing~\cite{Orzel2001a}, collapses and 
revivals~\cite{Greiner2002b}, 
and a number of 
superfluid properties of Bose gases~\cite{Cataliotti2001a,Burger2001a}.
Importantly, the Bose-Einstein condensates in an optical lattice realize
the Bose-Hubbard model~\cite{Jaksch1998a} and can be used to study
the quantum phase transition from the superfluid into a
Mott-insulator state~\cite{Greiner2002a,Jaksch1998a,vanOosten2001a}. 

Apart from these phenomena,
sound propagation is also possible in a Bose-Einstein condensate
in an optical lattice.
There exists several theoretical calculations for the 
sound velocity of a Bose-Einstein condensate in
a one-dimensional optical 
lattice~\cite{BergSorensen1998a,Javanainen1999a,vanOosten2001a,Kraemer2002a,
Machholm2003a,Martikainen2003a}. 
All these theories for the sound mode are based on the idea that
the condensate wave function in site $n$ is, in tight-binding approximation,
essentially one dimensional
and has the form $\Phi_n(x,y,t)=\psi(x,y)\sqrt{N_n(t)}e^{i\nu_n(t)}$.
This implies that the only dynamical variables of these
theories are the total number of atoms $N_n(t)$ 
and the global phase $\nu_n(t)$ in every site, but there is no
time dependence in the wave function describing the transverse directions.
Physically, it is not {\it a priori} clear that this is a valid
assumption because
in the presence of
repulsive interactions between the atoms the profile $\psi(x,y)$
does not remain constant as the number of particles in the
site is changed. For example, increasing the number of atoms from
its equilibrium value causes the condensate to expand whereas
a reduction of the number of atoms causes the condensate to contract.
Therefore, number fluctuations in every site
of the lattice are coupled
to collective modes of the condensate in each site.
In a one-dimensional optical lattice where the lattice runs
in the longitudinal direction (the $z$-direction), the coupling 
predominantly excites the breathing mode in the 
transverse direction (the $xy$-plane). These 
transverse modes were studied
previously in Ref.~\cite{Martikainen2003a}, in the 
approximation that the coupling with longitudinal sound
can be neglected.
While a coupling definitely exists between
the sound mode and the transverse breathing mode, a coupling 
between the transverse quadrupole modes and the
sound mode does not exist because these modes are orthogonal.
Therefore, the results in Ref.~\cite{Martikainen2003a} 
for the quadrupole modes are not affected by the
coupling with the sound mode.

In this paper we go beyond the approximation
used in 
Refs.~\cite{BergSorensen1998a,Javanainen1999a,vanOosten2001a,Kraemer2002a,
Machholm2003a,Martikainen2003a},
by providing a unified theory of longitudinal sound and the transverse
breathing mode in a one-dimensional optical lattice.
Using a variational approach 
we obtain  the magnitude of the change
to the sound velocity when the coupling with the transverse 
breathing mode is taken into account. It turns out that the relative 
shift in the sound velocity from the uncoupled result
approaches a constant $\sqrt{3/4}-1\approx -0.13$ as the strength
of the interactions is increased, i.e.,
in the Thomas-Fermi limit. Therefore, the shift is rather large
and should be taken into account when quantitative 
results are required. Likewise the shift in the effective
mass of the transverse breathing mode from the result
in Ref.~\cite{Martikainen2003a} also turns out to be large
and should also be taken into account.

Intriguingly, our approach leads to a very simple theory
of Josephson oscillations coupled to just one transverse
degree of freedom. Such a theory has some
analogies with the theories describing damped relative phase
dynamics of weakly-coupled condensates~\cite{Meier2001a,Zapata2003a}.
In these theories the damping of the relative
phase is caused by the coupling into quasiparticles. Our
minimal theory can be used to study the phase
dynamics of the Bose-Einstein condensate in a one dimensional
optical lattice and with considerable modifications even the
(quasi) irreversible damped phase dynamics.

This paper is organized as follows. In Sec.~\ref{sec:Ansatz},
we outline the theoretical foundation of our theory and explain
the variational ansatz we use.
In Sec.~\ref{sec:eqsandmodes} we write down the equations of
motion and solve for the eigenmodes. We then proceed
to dicuss, in subsections~\ref{sec:sound} and~\ref{sec:breath},
the dispersion relations for the sound and
the transverse breathing mode separately.
Previous experimental
studies of the condensate sound mode~\cite{Andrews1997b,Andrews1998a}
tracked the time evolution of a density dip.
Therefore in Sec.~\ref{sec:dipevolution}, we apply our theory to solve 
numerically the time evolution of such a dip in the 
condensate density in an optical lattice. 
In addition, in Sec.~\ref{sec:dipevolution}
we also solve  the time evolution
of the system that was prepared with a localized breathing-mode
disturbance.
%We find that the numerical result for the propagation velocity of
%the density disturbance, is in agreement with our analytical result.
We conclude 
with a brief discussion of our results in Sec.~\ref{sec:conclusions}.

\section{Variational ansatz and Lagrangian}
\label{sec:Ansatz}
The theory we use is similar to one presented 
elsewhere~\cite{Martikainen2003a}. In this paper the 
only difference
with the theory presented in Ref.~\cite{Martikainen2003a} is
the use of the grand-canonical Hamiltonian, i.e., inclusion
of the chemical potential term $-\mu\hat{N}$ into the
energy functional. Due to the presence of the global phase factors
and the atom-number fluctuations in different sites, this
turns out to be more convenient for our purposes.

We consider a Bose-Einstein condensate
trapped by a harmonic trap with  a radial trapping frequency $\omega_r$.
The longitudinal trapping frequency $\omega_z\ll \omega_r$ is assumed
to be so small as to be irrelevant.
The Bose-Einstein condensate also experiences a one-dimensional optical lattice
in the longitudinal direction and this lattice
splits the condensate into a stack of weakly-coupled 
two-dimensional condensates.
Furthermore, we use trap units, i.e., the unit of energy is $\hbar\omega_r$,
the unit of time is $1/\omega_r$,
and the unit of length is $l_r=\sqrt{\hbar/m\omega_r}$, where
$m$ is the atomic mass. The energy functional for the stack
of two-dimensional condensates is then
\begin{eqnarray}
\label{2D_Hamiltonian}
E\left[\Phi^*,\Phi\right]
&=&\sum_n \int d^2r\left\{-\frac{1}{2}\Phi_n^*(x,y)\nabla^2\Phi_n(x,y)
+\left[\frac{1}{2}\left(x^2+y^2\right)+
\right.\right.\\
&+&\left.\left.\frac{U_{2D}}{2}|\Phi_n(x,y)|^2-\mu
\right]|\Phi_n(x,y)|^2
-J\sum_{<n,m>}\int d^2r \Phi^*_m(x,y)\Phi_n(x,y)\right\},\nonumber
\end{eqnarray}
where the lattice sites are labeled by $n$ and
$\langle n,m \rangle$ indicates nearest neighbours. 
This energy functional is characterized by two tunable
paremeters: the strength of the interaction
$U_{2D}$ in every two-dimensional Bose-Einstein condensate
and the strength $J$ 
of the Josephson coupling between the condensates in the
neighbouring sites. In terms
of the scattering length $a$, the trapping frequency 
$\omega_L$ in every site due to the optical lattice, and the
characteristic length scale
$l_L=\sqrt{\hbar/m\omega_L}$, the interaction strength
is given by
\beq
U_{2D}=4\sqrt{\frac{\pi}{2}}
\left(\frac{a}{l_L}\right).
\enq
Moreover,
approximating the lattice potential near its maximum by an upside-down 
parabolic potential we obtain for the Josephson coupling
\beq
J=\frac{1}{8\pi^2}\left(\frac{\omega_L}{\omega_r}\right)^2
\left(\frac{\lambda}{l_r}\right)^2\left[\frac{\pi^2}{4}-1\right]
e^{-\left(\lambda/4\,l_L\right)^2},
\enq
where $\lambda$
is the wavelength of the laser beams creating
the optical lattice.

We use a variational approach
to study the sound and the transverse breathing modes 
of the condensate. For
this purpose we use the gaussian ansatz
\beq
\Phi_n(x,y,t)=
\sqrt{\frac{N\left(B_0+\epsilon'(t)\right)\left(1+\delta_n(t)\right)}{\pi}}
\,\exp
\left(-\frac{\left(B_0+\epsilon_n(t)\right)\left(x^2+y^2\right)}{2}
+i\nu_n(t)\right)
\label{eq:ansatz}
\enq
for the two-dimensional wave function at site $n$. The variational parameters
$\epsilon_n=\epsilon_n'+i\epsilon_n''$ are complex and describe
the amplitude of the breathing mode, whereas $\delta_n$ and $\nu_n$
are the relative 
number fluctuation and the global phase at site $n$, respectively.
Furthermore, $N$ is the equilibrium number of atoms at the site and $B_0$ gives
the equilibrium size of the two-dimensional condensate. 
The latter is obtained
by minimizing the equilibrium energy functional. 

Using this ansatz we can calculate the energy functional.
We find that in equilibrium the condensate
size parameter $B_0$ is given by
\beq
B_0=\frac{1}{\sqrt{1+2U}}\,,
\enq
where, for convenience, we defined the strength of the interaction as
\beq
U=\frac{N}{\sqrt{2\pi}}\left(\frac{a}{l_L}\right).
\enq
In addition, the chemical potential is given by
\beq
\mu=\frac{3}{2B_0}-\frac{B_0}{2}-2J,
\enq
because this choice for the chemical potential removes
the terms linear in $\delta_n$ from the energy functional.

As we are interested in the collective modes we must
expand the Lagrangian $L=T-E$ up to second order in the variational
parameters.  Due to the technical simplicity of
the number and global phase fluctuations we choose, 
for the time being, to treat them exactly and only expand
in $\epsilon_n$.
It then turns out that the contribution from the time-derivative
term in the Lagrangian is
\begin{eqnarray}
T/N&=&\frac{i}{2}\sum_n\int d^2r
\left[\Phi_n^*\frac{\partial\Phi_n}{\partial t}-
\Phi_n\frac{\partial\Phi_n^*}{\partial t}\right]=
\nonumber\\
&=&\sum_n \left(1+\delta_n\right)\left[-\dot{\nu}_n+
\frac{\dot{\epsilon}_n''}{2B_0}
\left(1-\frac{\epsilon_n'}{B_0}\right)\right]
\end{eqnarray}
whereas the energy functional becomes
\begin{eqnarray}
E/N&=&H_J+\sum_n \left[\left(\frac{1+\delta_n}{2B_0}\right)
\left(1-\frac{\epsilon_n'}{B_0}\right)
+\left(\frac{\epsilon_n'}{B_0}\right)^2+
\frac{B_0}{2}\left(1+\delta_n\right)\left(1+\frac{\epsilon_n'}{B_0}
+\left(\frac{\epsilon_n''}{B_0}\right)^2\right)
\right.\nonumber\\
&+&\left.UB_0\left(1+\delta_n\right)^2\left(1+\frac{\epsilon_n'}{B_0}\right)
-\mu N\left(1+\delta_n\right)\right].
\label{eq:energy}
\end{eqnarray}
In Eq.~(\ref{eq:energy}) we have explicitly split off
the contribution from the Josephson coupling
\beq
H_J=-J\sum_{<n,m>}\left[\cos\left(\nu_n-\nu_m\right) I_{nm}'
-\sin\left(\nu_n-\nu_m\right) I_{nm}''
\right],
\enq
where $I_{mn}=\int d^2r \Phi_m^*(x,y)\Phi_n(x,y)$ is the overlap integral. 
Expanding the overlap integral 
up to second order, now also in
the number and global phase fluctuations, we readily obtain
\begin{eqnarray}
H_J&=&\frac{J}{8B_0^2}\sum_{\langle n,m\rangle}
\left[\left(\epsilon_n'-\epsilon_m'\right)^2+
2\left(\epsilon_n''-\epsilon_m''\right)^2
+4B_0^2\left(\nu_n-\nu_m\right)^2
\right.\nonumber\\
&+&\left.B_0^2\left(\delta_n-\delta_m\right)^2
-4B_0\left(\nu_n-\nu_m\right)\left(\epsilon_n''-\epsilon_m''\right)
\right].
\end{eqnarray}
From this expression it is clear how the global phase
fluctuations are coupled to the transverse breathing mode.
Technically this is due to the imaginary part of the overlap integral
which has a contribution linear in $\left(\epsilon_n''-\epsilon_m''\right)$.
Such a contribution does not exist, for example, for the quadrupole modes
which preserve the condensate volume. Therefore, quadrupole modes will
not be affected by the presence of the sound mode.

In the above result for the Josephson energy
$H_J$ we expanded also in terms of the
number $\delta_n$ and global phase $\nu_n$ fluctuations.
In the problem we are focusing on here these fluctuation
are small and the above procedure is justified. 
In principle, however,
number and global phase fluctuations can also be included exactly.
By including them exactly we can also capture the
physics of modulational and dynamical instabilities
in an optical lattice~\cite{Wu2001a,Konotop2002a,Baizakov2002a}, but
this is outside the scope of the present paper.

\section{Equations of motion and the eigenmodes}
\label{sec:eqsandmodes}
The results obtained in the previous section
enable us to derive the linearized Euler-Lagrange equations of motion
for the variational parameters. They read
\beq
\frac{\partial \epsilon_n'}{\partial t}=2B_0\epsilon_n''
+J\sum_{\langle n,m\rangle}\left(\epsilon_n''-\epsilon_m''\right),
\label{eq:er}
\enq
\beq
\frac{\partial \epsilon_n''}{\partial t}=-\frac{2}{B_0}\epsilon_n'
-2UB_0^2\delta_n-J\sum_{\langle n,m\rangle}
\left(\epsilon_n'-\epsilon_m'\right),
\enq
\beq
\frac{\partial \nu_n}{\partial t}=-\left(1+3U\right)\epsilon_n'
-3B_0U \delta_n-\frac{J}{2}\sum_{\langle n,m\rangle}
\left[\left(\delta_n-\delta_m\right)+\left(\frac{\epsilon_n'-\epsilon_m'}{B_0}
\right)\right],
\enq
and
\beq
\frac{\partial \delta_n}{\partial t}=2J\sum_{\langle n,m\rangle}
\left[\left(\nu_n-\nu_m\right)-
\frac{\left(\epsilon_{n}''-\epsilon_m''\right)}{2B_0}\right].
\label{eq:delta}
\enq
These equations of motion can also be written in a more formal way as
\beq
\frac{\partial \epsilon_n'}{\partial t}=2B_0^2
\left(\frac{\partial E}{\partial\epsilon_n''}\right)+
B_0\left(\frac{\partial E}{\partial\nu_n}\right),
\label{eq:ereq}
\enq
\beq
\frac{\partial \epsilon_n''}{\partial t}=-2B_0^2
\left(\frac{\partial E}{\partial\epsilon_n'}\right),
\enq
\beq
\frac{\partial \nu_n}{\partial t}=
-\left(\frac{\partial E}{\partial \delta_n}\right)
-B_0\left(\frac{\partial E}{\partial\epsilon_n'}\right),
\label{eq:phaseeq}
\enq
and
\beq
\frac{\partial \delta_n}{\partial t}=\frac{\partial E}{\partial\nu_n}.
\enq
While the first form of the equations is eventually needed in the 
actual calculations, the second way of writing the equations 
provides some additional insight.
In particular, by inspecting Eqs.~(\ref{eq:ereq})-(\ref{eq:phaseeq})
it becomes clear how the global phases
influence the behaviour of the breathing mode through the term
$B_0\left(\partial E/\partial\nu_n\right)$ in
Eq.~(\ref{eq:ereq}) and how the breathing mode
influences the dynamics of the global phase through
the term $-B_0\left(\partial E/\partial\epsilon_n'\right)$
in Eq.~(\ref{eq:phaseeq}). Without these terms
we could treat the condensate density fluctuations
independently from the transverse breathing mode.

The  four first-order differential equations
Eqs.~(\ref{eq:er})-(\ref{eq:delta}) 
for the variational parameters
can  be cast into two coupled second-order differential equations 
for $\epsilon_n'$ and $\delta_n$. We are looking for solutions
of the type $\epsilon_n'=\epsilon_k'(t)\sin\left(nk\lambda/2\right)$
and $\delta_n=\delta_k(t)\sin\left(nk\lambda/2\right)$. By inserting these
into the equations of motion we obtain
%\beq
%\frac{d}{dt^2}\left(\begin{array}{c}
%\epsilon_k' \\ \delta_k\end{array}\right)
%=\hat{\Omega}^2 \left(\begin{array}{c}
%\epsilon_k' \\ \delta_k\end{array}\right)
%\enq
\beq
\frac{d^2\epsilon_k'}{dt^2}=-\left(2B_0+2J(k)\right)
\left(\frac{2}{B_0}+2J(k)\right)\epsilon_k'+
\left(B_0^2-1\right)\left(2B_0+2J(k)\right)\delta_k
\label{eq:erfluc}
\enq
and
\beq
\frac{d^2\delta_k}{dt^2}=4J(k)\left(B_0-\frac{1}{B_0}-J(k)\right)
\delta_k-\frac{2J(k)}{B_0}\left(3-3B_0^2-\frac{2}{B_0}
\right)\epsilon_k',
\label{eq:numfluc}
\enq
where $J(k)=J\left(1-\cos(k\lambda/2)\right)$.
The two eigensolutions of these equations correspond to the
longitudinal sound mode and the transverse breathing mode.
These equations of motion and their solutions are the
main results of this paper.

\begin{figure}
\includegraphics[width=\columnwidth]{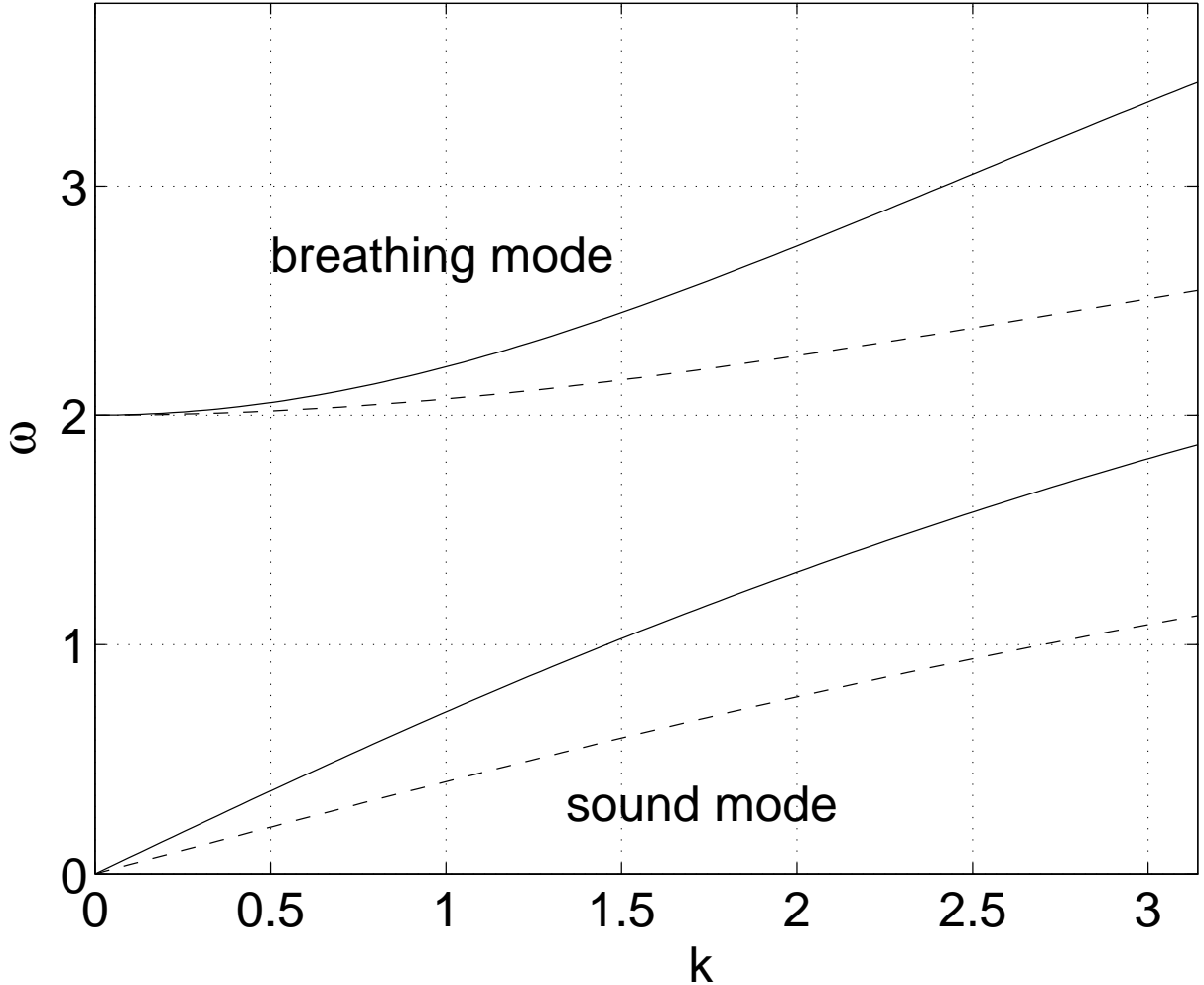}
\caption[]{Eigenmode frequencies of the Eqs.~(\ref{eq:erfluc})
and~(\ref{eq:numfluc}) as a function of momentum, 
when $J=0.1$, $U=100$ (solid line), and
$U=10$ (dashed line).
\label{fig:twobranches}}
\end{figure}

In Fig.~\ref{fig:twobranches} we show a typical behaviour 
of both the sound mode and the transverse breathing-mode
frequency
as a function of momentum $k$ for
two different values of the interaction strength.
This figure demonstrates how the small $k$ behaviour of the sound modes
is linear whereas that of the breathing mode is quadratic.
It also shows that, 
both modes become stiffer as the strength of the interactions
is increased. 
We devote the following two subsections
to the discussion of both modes separately.

\subsection{The longitudinal sound mode}
\label{sec:sound}
Theories of bosons in a lattice that reduce the problem
to that of a global phase and an atom number in each site, can be
used to solve the dispersion relation $\omega_S(k)$
of the sound mode. We refer to these theories as
phase-only theories. At small values of the momenta,
the dispersion relation of the sound mode 
is linear in momentum, i.e., $\omega_S(k)=c_{0}k$.
By using the ansatz~(\ref{eq:ansatz}), but removing the
breathing modes, we find the phase-only sound velocity 
\beq
c_{0}=\lambda\sqrt{JB_0U}.
\label{eq:phaseonlysound}
\enq
This agrees with the results of 
Refs.~\cite{BergSorensen1998a,Javanainen1999a,vanOosten2001a,Kraemer2002a,
Machholm2003a,Martikainen2003a}.

However, this result is changed when the transverse
breathing mode is taken into account. Using the unified description
of sound and breathing modes that we
have presented in this paper, we obtain a different 
sound velocity, namely
\beq
c=\lambda\sqrt{\frac{J}{8}\left(\frac{3}{B_0}
-2B_0-B_0^3\right)}.
\label{eq:exactsound}
\enq
To obtain this simple expression we ignored contributions
proportional to $J^2$ as these are in general very small.

\begin{figure}
\includegraphics[width=\columnwidth]{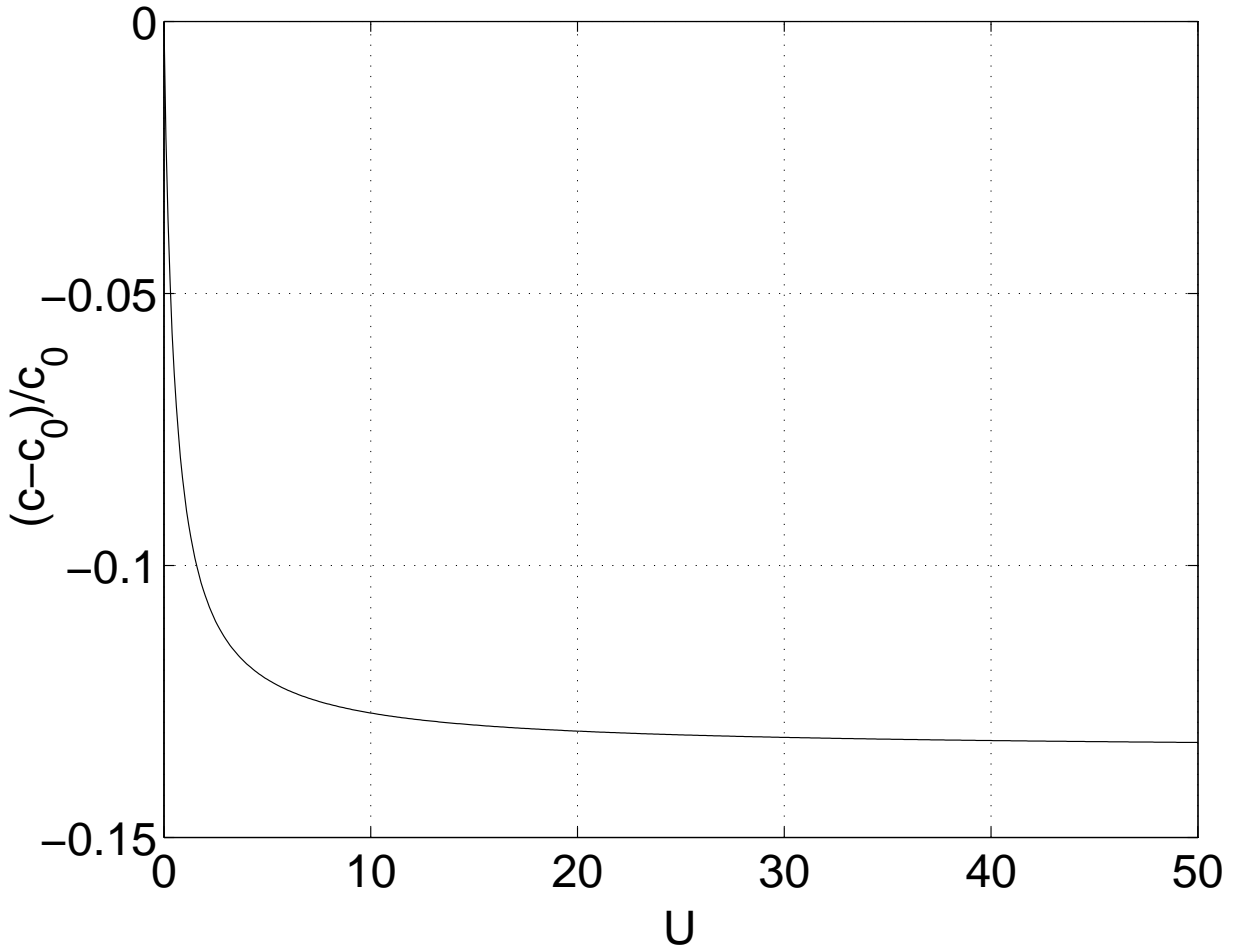}
\caption[]{The relative shift in the sound velocity $c$ from 
the phase-only result in Eq.~(\ref{eq:phaseonlysound}) 
as a function of the interaction strength.
\label{fig:soundshift}}
\end{figure}

In Fig.~(\ref{fig:soundshift}) we show the relative shift
$c/c_0-1$ in the sound velocity from 
the phase-only theory.
In the non-interacting limit 
the two results coincide, but the shift becomes more
pronounced with increasing interaction strength.
In particular, the real
sound velocity is smaller than the one predicted by
the phase-only theories. Asymptotically we obtain
in the limit of strong interactions 
$\lim_{U\rightarrow\infty}\left(c/c_0-1\right)=
\sqrt{3/4}-1\approx -0.13$. While this shift is not enormous,
it is still large enough to be kept in mind when
quantitative results are required.
Intuitively, the reduction of the real
sound velocity due to the transverse breathing mode
is expected, since
the transverse degree of freedom makes the condensate
less stiff. Rather than all the energy 
of the density disturbance being pushed
forward in the longitudinal
direction, some of the excess energy is lost in exciting the transverse
degree of freedom.

It is interesting to observe that while a ``pure''
breathing mode, unaffected by the global phase dynamics,
is possible when all the sites are breathing in phase, a ``pure''
sound mode is never possible. The sound mode relies
on global phase differences and number fluctuations between sites,
and these will inevitably couple into the transverse breathing mode.

\subsection{The transverse breathing mode}
\label{sec:breath}
At small values of momenta 
the transverse breathing mode in the absence
of sound mode has a dispersion relation~\cite{Martikainen2003a}
\beq
\omega_0=2+\frac{J\lambda^2}{8}\left(B_0+\frac{2}{B_0}\right)k^2
\label{eq:breathingonly}.
\enq
However, in line with our previous result for the sound mode,
Eqs.~(\ref{eq:erfluc}) and~(\ref{eq:numfluc}) predict
a different behaviour 
for the term proportional to $k^2$. Namely, we have
\beq
\omega_B=2+\frac{J\lambda^2}{16}
\left(B_0+\frac{5}{2B_0}+\frac{B_0^2}{2}\right)
k^2,
\enq
where we again ignored contributions proportional to $J^2$.

\begin{figure}
\includegraphics[width=\columnwidth]{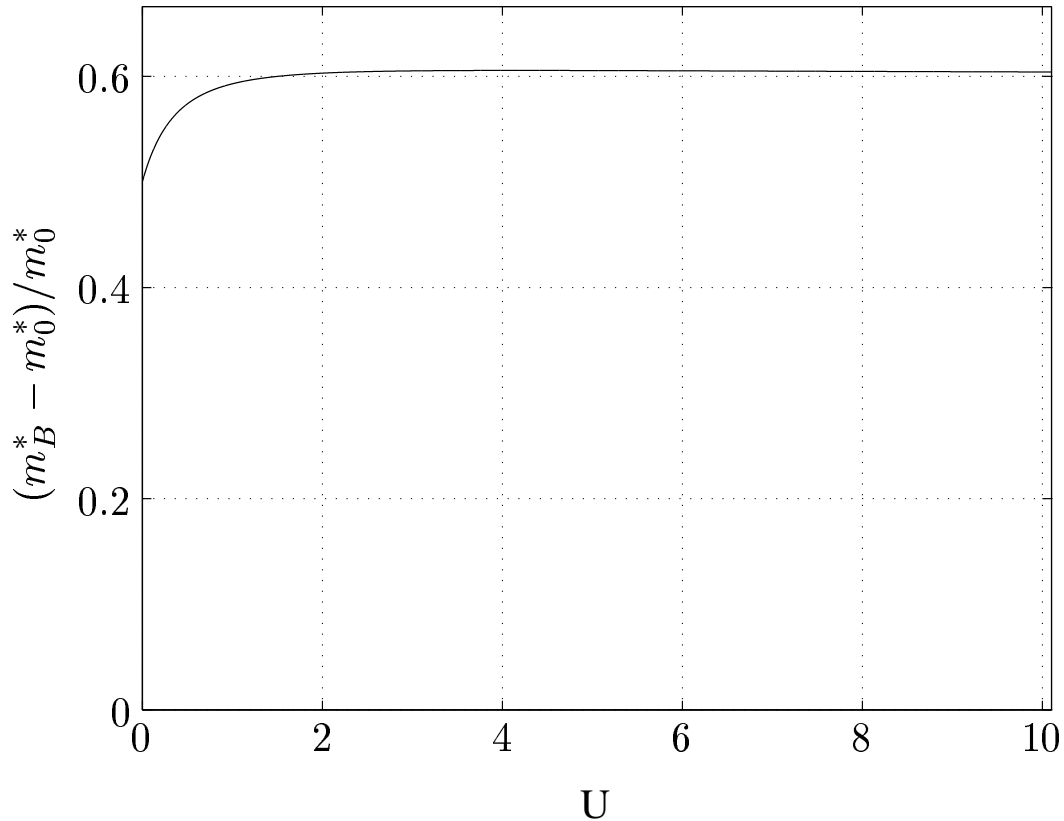}
\caption[]{The relative shift in 
the effective mass $m_B^*$  of the transverse breathing mode 
from the uncoupled result in Eq.~(\ref{eq:breathingonly}) 
as a function of the interaction strength.
\label{fig:breathingshift}}
\end{figure}

The breathing-mode dispersion relations can be given 
in terms of an effective masses $m_0^*$ and $m_B^*$ 
for the transverse breathing mode as
$\omega_0=2+k^2/2m_0^*$ and $\omega_B=2+k^2/2m_B^*$.
In Fig.~(\ref{fig:breathingshift}) we show the relative shift
$m_B^*/m_0^*-1$ in the effective mass of the transverse
breathing mode. The relative shift in the effective mass 
is always quite large and quickly approaches an asymptotic value $3/5$
as the strength of the interaction increases.
This shift, therefore, must be included
in making quantitive predictions for upcoming experiments.

\section{Time evolution of the coupled system}
\label{sec:dipevolution}
Earlier in this paper we solved the equations of motion for the
sound and the transverse breathing mode by assuming
plane-wave solutions. This enabled us to obtain 
analytic solutions for the dispersion relations,
but our theory can be used to solve also more 
complicated problems. In this section we demonstrate this
by solving the coupled 
dynamics of the density fluctuations and the transverse breathing mode,
when the initial state of the condensate has a density
dip. This problem is interesting since the earlier
experiments on a condensate sound mode first created
a density dip and then tracked the evolution of the condensate
density~\cite{Andrews1997b,Andrews1998a}.
In this case 
calculating the time evolution of the coupled system
of the sound mode and the transverse breathing mode 
is too complicated to be attacked analytically,
but the problem can be readily tackled numerically. 

In Fig.~\ref{fig:dipevolution1} we show an example of 
the typical time evolution. In this figure we prepare the system
with a Gaussian density disturbance
$\delta_n(t=0)=-0.1\exp\left[-\left(n/10\right)^2\right]$ and
then let it evolve. The density minimum splits into
two parts propagating into opposite
directions. The propagation velocity of
the disturbance is in agreement
with our result for the sound velocity 
in Eq.~(\ref{eq:exactsound}), but is different from the phase-only result
in Eq.~(\ref{eq:phaseonlysound}). The excess energy
of the density dip excites the transverse breathing mode that 
remains well localized in the center of the lattice
and only slowly spreads out further. This slow
spreading of the breathing mode is hardly visible at the
relatively short timescale of the figure.

\begin{figure}
\caption[]{The time evolution of (a) the density disturbance
$\delta_n(t)$
and (b) the transverse breathing-mode amplitude $|\epsilon_n(t)|^2$.
We used $J=0.1$, $U=100$, and the number of sites was $101$.
The initial state had a Gausssian density disturbance
$\delta_n(t=0)=-0.1\exp\left[-\left(n/10\right)^2\right]$.
In the figure dark color indicates the disturbance.
It can be seen how the initial density dip
splits into two dips propagating into opposite directions whereas
at this timescale the breathing mode remains well localized
around the location of the initial density disturbance.
\label{fig:dipevolution1}}
\end{figure}

In Fig.~\ref{fig:breathevolution} we demonstrate the ``inverse''
problem of an initial state that has a localized
transverse breathing-mode disturbance and a homogeneous
density distribution. The time evolution of the density
distribution is then more complicated, but the time evolution of
the breathing-mode amplitude is similar to
that in Fig.~\ref{fig:dipevolution1}. In Fig.~\ref{fig:breathevolution}
the fractional change of the condensate size was $-25\,\%$
in the center of the lattice and such a deformation is 
quite large. Despite this strong deformation, the magnitude of the
density disturbance remains small, below $1\,\%$. Therefore,
under these conditions
it would be difficult to image the density disturbance
experimentally.
 
\begin{figure}
\caption[]{The time evolution of (a) the density disturbance
$\delta_n(t)$
and (b) the transverse breathing-mode amplitude $|\epsilon_n(t)|^2$.
We used $J=0.1$, $U=100$, and the number of sites was $101$.
The initial state had a Gausssian transverse breathing mode disturbance
$\epsilon_n(t=0)=B_0/2\,\exp\left[-\left(n/10\right)^2\right]$.
This choice corresponds to the fractional change of
$-25\,\%$
of the condensate size in the center of the lattice.
In (a) dark color indicates a region of low density.
In (b) it would seem that the breathing-mode amplitude is
initially zero. This is a result of plotting the amplitude squared
$|\epsilon_n(t)|^2$ as opposed to plotting just the real part 
$\epsilon_n'(t)$. Typically the imaginary part of the breathing mode 
amplitude has a larger magnitude and therefore dominates in the
amplitude squared plot.
\label{fig:breathevolution}}
\end{figure}

\section{Summary and conclusions}
\label{sec:conclusions}
We have presented a unified theory of the sound propagation
and the transverse
breathing mode of a Bose-Einstein condensate 
in a one-dimensional optical lattice. 
Using a variational ansatz
we calculated the dispersion relations of both modes
and found out that the dispersion relations are quite
strongly modified by the coupling between the sound and
the transverse breathing mode. These changes are
large enough that they should be included when making
quantitative predictions for experiments.
In principle, the sound mode is not only coupled to the
transverse breathing mode, but also to modes with
higher energy that have the same symmetry. 
In this paper we have ignored such higher-order effects, since
the overlap with the transverse breathing mode 
is the largest and therefore
the coupling to this mode dominates. 
 
In principle, the theories studying two weakly-coupled
condensates~\cite{Smerzi1997a,Zapata1998a,Raghavan1999a} 
would also be influenced by the mechanism
we have discussed in this paper. This makes the study of
the phase dynamics 
of the weakly-coupled condensates in the presence of 
the transverse degrees of freedom an interesting topic 
for further research. 

When the sound mode in an ordinary
elongated Bose-Einstein condensate is excited, by using for example
a blue-detuned laser beam as in the experiments
of Andrews {\it et al.}~\cite{Andrews1997b,Andrews1998a}, 
the transverse profile of the
condensate is modified close to the laser beam. After removing 
the laser beam, 
the density dip starts propagating with the sound velocity.
However, we have seen that in principle in addition to this also a surface
disturbance starts propagating in the condensate. This is
analogous to the problem we have studied in this paper.
In contrast with the Bose-Einstein condensate in an optical lattice, 
in a cigar shaped three-dimensional Bose-Einstein condensate
the influence of the transverse degrees of freedom
is expected to be small, and theories
ignoring them~\cite{Zaremba1998a,Kavoulakis1998a,Stringari1998a} 
are indeed in agreement 
with the experiments~\cite{Andrews1997b,Andrews1998a,StamperKurn1999a}.

Our theory can also be
used to study how the transverse breathing mode
excites density modulations of the condensate 
in an optical lattice. In this paper we gave one numerical example 
along these lines, but when the transverse breathing-mode
amplitude is very large or strongly modulated, we expect
that nonlinearities will play an important role. Under such
conditions instabilities might arise, revealing new
possibilities for studying nonlinear matter wave dynamics.

\begin{acknowledgments}
This work is supported by the Stichting voor Fundamenteel Onderzoek der 
Materie (FOM) and by the Nederlandse Organisatie voor 
Wetenschaplijk Onderzoek (NWO).
\end{acknowledgments}

%\bibliographystyle{apsrev}
%\bibliography{bibli}

\end{document}